\documentstyle[12pt]{article} 

\begin{document}
 
\newcommand{\beq}{\begin{equation}}
\newcommand{\eeq}{\end{equation}}
\newcommand{\barr}{\begin{eqnarray}}
\newcommand{\earr}{\end{eqnarray}}
\newcommand{\andy}[1]{ }

\def\ask{\marginpar{?? ask:  \hfill}}
\def\fin{\marginpar{fill in ... \hfill}}
\def\note{\marginpar{note \hfill}}
\def\check{\marginpar{check \hfill}}
\def\discuss{\marginpar{discuss \hfill}}


\author{{\bf Fran\c{c}ois Goy}  \\
      \\     
        Dipartimento di Fisica,
Universit\^^ {a} di Bari \\
Via G. Amendola 173 \\
I-70126  Bari, Italy \\ 
E-mail: GOY@AXPBA1.BA.INFN.IT}
        
\title{{\bf Aberration and the Question of Equivalence of some Ether 
Theories to Special Relativity }}
 
\date{December 7,95}
 
\maketitle
 
\begin{abstract}
In the last two decades, theories explaining the same experiments as well as 
special relativity does, were developed by using different synchronization 
procedures. All of them are ether-like theories. Most authors believe these 
theories to be equivalent to special relativity, but no general proof was ever
brought. By means of a Gedankenexperiment on light aberration, we produce 
strong evidence that this is the case for experiments made in inertial systems.
\\
\\
{\bf Keywords}: special relativity, ether, velocity of light, stellar 
aberration, synchronization.
\end{abstract}

\setcounter{equation}{0}

\section{Introduction}
In his famous article of 1905, Einstein\cite{eins:05a} gave a {\em definition} 
of time in a
 point B, in terms of the time of a distant point A in 
the same inertial system, by postulating that the ``time'' which light needs
to go from A to B is the same as the ``time'' needed 
to go from B to A. He then
showed that this synchronization procedure is consistent with his relativity 
principle. 

More recently, Mansouri and Sexl\cite{mase:77a} have constructed a set of
transformations between inertial systems, based on other synchronizations,
which depend on a parameter $\varepsilon$. They showed that only the value 
of $\varepsilon$ given by the Einstein's 
synchronization procedure is compatible with the principle of relativity and
gives a one-way velocity of light constant in all inertial frames. 
In particular, the 
choice of the so-called absolute synchronization leads to an ether-like theory
(Theory of Inertial Transformations(TIT))
which has transformations different from the Lorentz-ones, maintains absolute 
simultaneity, and has a one-way velocity of light different from c. It was 
claimed by the authors, that this ether-like theory is kinematically 
equivalent 
to the Special Relativity Theory(SRT). This implies that it is impossible to
 measure the velocity 
of an inertial frame relative to the ether, or equivalently, that the one-way 
velocity of light is a convention, or that the synchronization procedure is 
 a matter of choice. This is sometimes called Poincar\'{e} principle, 
but it is a conjecture that was really never proved.

Sj\"{o}din\cite{sjod:79a} developed waves and Maxwell equations for all 
synchronization procedures. He stated that we can only measure the ``absolute'' 
velocity  of inertial systems by means of tachyons or waves travelling 
through 
the ether at a velocity different from c. Since such things have never been 
observed and probably do not exist in nature, this idea has no practical 
consequences. 

Selleri\cite{sell:95a} developed the dynamical part of the TIT and the general
transformations between inertial systems. He gave an argument based on the
Sagnac effect claiming the logical inequivalence of the SRT and TIT. In fact, 
if there were only one different prediction between the TIT and the SRT, it 
would imply
that the one-way velocity of light is measurable independently of conventions. 

Aberration of light is a phenomenon in which only the one-way velocity of 
light
come into play and in which, apparently, no clocks are used. Thus we think it 
is a
good test for the equivalence of the SRT and the TIT. Sj\"{o}din and 
Podlaha\cite{sjpo:81a} wrote an article on the subject but their earth was at 
rest always in the same inertial frame, thus an idealized solar system
very different from the real one.

We develop here a Gedankenexperiment using aberration of light with two
different inertial frames and conclude that the SRT leads
to the same results as the TIT.

\section{Synchronization and measurement of velocities}
Let $K$ be an inertial frame having velocity \mbox{\boldmath $v$} relative to
the fundamental frame $K_0$ along the $x_0$-axis in positive direction.
Along the $x$-axis of $K$, there are two points A and B with
$x_{{\sc a}} < x_{{\sc b}}$ and 
$\mid$\mbox{\boldmath $x_{{\sc b}}$} - 
\mbox{\boldmath $x_{{\sc a}}$}$\mid = d$. In
A there is one clock and in B there are two clocks. 

We synchronize the first clock in B with the one in A
by using Einstein's procedure. It means that a light ray is sent from A 
 at time $t_{{\sc a}}$ reflected in B at time $\tilde{t}_{{\sc b}}$
(where $\tilde{ }$ stands for Einstein's synchronization) and comes back to
A at time $t_{{\sc a}}^{*}$. The {\em definition} of 
$\tilde{t}_{{\sc b}}$ is 
\andy{joelle}
\beq
\tilde{t}_{{\sc b}} = \frac{t_{{\sc a}}^{*} + t_{{\sc a}}}{2}
\label{eq:joelle}
\eeq
One can easily verify that this definition is based on the assumption that
the velocity of light is the same from A to B as from 
B to A.
If an object on which no forces act leaves A at time $t_1$ and 
reaches B at time $\tilde{t}_2$, its velocity $\tilde{v}_1$ is
\andy{murielle}
\beq
\tilde{v}_1 = \frac{d}{\tilde{t}_2 - t_1}
\label{eq:murielle}
\eeq

Then, we synchronize the second clock in B with a procedure of
``absolute'' synchronization, which takes account of the fact that the
velocity of light $c_{\sc ab}$ from A to B is different from
 the velocity $c_{\sc ba}$ from B to A. The TIT gives
(see Ref. \cite{sell:95a})
\andy{lilas}
\beq
c_{\sc ab} = \frac{c}{1+\beta};\;\;\;\;\;
c_{\sc ba} = \frac{c}{1-\beta}
\label{eq:lilas}
\eeq
where $c$ is the two-way velocity of light and $\beta$ stands for 
$\frac{v}{c}$.
A light ray is sent from A at time $t_{{\sc a}}$, reflected in 
B at time $t_{{\sc b}}$ and comes back to A at time 
$t_{{\sc a}}^{*}$. Using (\ref{eq:lilas}), we have:
\andy{judith}
\beq
d\;=\;c_{{\sc ab}}(t_{{\sc b}}-t_{{\sc a}})\; =\;
\frac{c(t_{{\sc b}}-t_{{\sc a}})}{1+\beta}\; =\;
c_{{\sc ba}}(t_{{\sc a}}^{*}-t_{{\sc b}})\; =\;
\frac{c(t_{{\sc a}}^{*}-t_{{\sc b}})}{1-\beta}
\label{eq:judith}
\eeq
Comparing the third and the fifth term of this equality, we obtain:
\andy{marie}
\beq
t_{{\sc b}} = \frac{t_{{\sc a}}^{*} + t_{{\sc a}}}{2} +
\frac{\beta(t_{{\sc a}}^{*} - t_{{\sc a}})}{2}
\label{eq:marie}
\eeq
In both theories, the two-way velocity of light is constant in all 
directions
so that:
\andy{anna}
\beq
t_{{\sc a}}^{*} - t_{{\sc a}} = \frac{2d}{c}
\label{eq:anna}
\eeq
Using (\ref{eq:joelle}) and (\ref{eq:anna}), Eq. (\ref{eq:marie}) becomes:
\andy{myriam}
\beq
t_{{\sc b}} = \tilde{t}_{{\sc b}} + \frac{\beta d}{c}
\label{eq:myriam}
\eeq
The velocity $v_1$ of the same object as before is measured. It leaves A
at time $t_1$ and reaches B at time $t_2$. Using
(\ref{eq:murielle}) and (\ref{eq:myriam}) with $t_{{\sc b}} = t_2$ and 
$t_{{\sc a}} = t_1$, we obtain:
\andy{genevieve}
\beq
v_1 = \frac{d}{t_2 - t_1} =
\frac{d}{\tilde{t}_2 - t_1 + \frac{\beta d}{c}} =
\frac{1}{\frac{1}{\tilde{v}_1} + \frac{\beta}{c}} =
\frac{\tilde{v}_1}{1+\beta\tilde{\beta}_1} 
\label{eq:genevieve}
\eeq
where $\tilde{\beta}_1$ stands for $\frac{\tilde{v}_1}{c}$.
In the same physical situation, an observer will not obtain the same 
numerical
value of the velocity if he uses different synchronization procedures, but
(\ref{eq:genevieve}) brings a connection between them. Note that the
results obtained here for the time $t_{{\sc b}}$ or the velocity 
$v_1$ can easily be generalized to the case of a vector $\vec{AB}$
making an angle $\psi$ with \mbox{\boldmath $v$}: at every place where
 $\beta$ appears it has to be replaced by $\beta\cos\psi$.

\section{Aberration}
Stellar aberration, discovered by Bradley in 1728, is an apparent motion
of all stars during the year along an ellipse, whose major
axis approaches 41" and is explained by the classical formula 
\andy{nina}
\beq 
\Delta\theta = \frac{\tilde{v}_1}{c}\sin\theta \;\;\;\;  
for\;\; \frac{\tilde{v}_1}{c}\ll 1
\label{eq:nina}
\eeq
where $\theta$ is the angle between the earth velocity and the light ray.
$\Delta\theta $ is the difference between the angles that light makes in two
reference frames moving at relative velocity $\tilde{v}_1$ 
(here the earth at 
different times of the year), and $c$ is the two-way speed of light.

In the SRT, the aberration is {\em only} a matter of motion {\em of 
the observer's} inertial frame relative 
to another inertial frame (see for example Ref. \cite{harw:74a}).
Considering an inertial system $K$ and another $K'$ moving with 
velocity \mbox{\boldmath $\tilde{v}_1$} relative to $K$ along the 
$x$-axis in positive direction, we have:
\andy{pascale}
\beq
\tan\tilde{\theta}' = R(\tilde{\beta}_1)\;\; 
                     \frac{\sin\theta}{\cos\theta + \tilde{\beta}_1}
\label{eq:pascale}
\eeq
where $\theta + \pi$ is the angle 
between the light velocity vector and the $x$-axis
, $\tilde{\theta'}+ \pi$ is the same quantity in $K'$ and 
$R(\tilde{\beta}_1)$
stands for $\sqrt{1-\tilde{\beta}_1^2}$. There is no 
tilda on $\theta$ because this angle does not depend at all on 
synchronisation, but there is one on $\tilde{\theta}'$ because this
angle is a function of $\tilde{v}_1$ in Eq. (\ref{eq:pascale}).

Note that
(\ref{eq:nina}) was derived from a classical equation similar to 
(\ref{eq:pascale}) but without the relativistic factor 
$R(\tilde{\beta}_1)$. Within the actual 
precision of measurement, this 
factor is not observable because the next term of $\Delta
\theta$ is of order $(\frac{\tilde{v}_1}{c})^3$ when $\theta=\pi/2$ 
and (\ref{eq:pascale}) reduces to (\ref{eq:nina}) for 
$\tilde{v}_1 \ll c$. 

Many textbooks and the original Einstein article explain aberration from
the relative motion of an observer and the {\em source}. It has been shown
that this explanation cannot be correct, because otherwise binary stars 
should 
present an aberration larger than 
other stars and they do not (see Refs.
\cite{eisn:67a,phip:89a,prmo:89a,hayd:93a}). 
The right relativistic explanation is the one given above.

In the TIT, the aberration equation depends on the ``absolute'' velocity 
of the observer. Let $K$ be an inertial frame moving with velocity 
\mbox{\boldmath $v$} 
relative to the privileged frame $K_0$ along the $x_0$-axis in positive
direction and another inertial frame $K'$ moving with velocity 
\mbox{\boldmath $v'$} along the $x_0$-axis in the positive direction. 
The inertial transformations in 2+1 dimensions between $K$
 and $K'$ (see Refs.
\cite{sell:95a}) are, with an obvious notation:
\andy{martine}
\begin{eqnarray}
x'& =& \frac{R(\beta)}{R(\beta')}\left[x-\frac{\left(\beta' - \beta\right)c}
{R^2(\beta)}\;\;t\right]\nonumber\\
y'& =& y \nonumber\\
t'& =& \frac{R(\beta')}{R(\beta)}\;\;t
\label{eq:martine}
\end{eqnarray}
where $R(\beta) = \sqrt{1-\beta^2}$ and $R(\beta') = \sqrt{1-\beta'^2}$.
Writing (\ref{eq:martine}) in differential form and dividing the space 
variables by the time variable, we obtain the transformations of velocity
components:
\andy{mona}
\begin{eqnarray}
u'_x & =& \frac{R^2(\beta)}{R^2(\beta')}
\left[u_x-\frac{\left(\beta' - \beta\right)c}
{R^2(\beta)}\right]\nonumber\\
u'_y & =& \frac{R(\beta)}{R(\beta')}\;\;u_y
\label{eq:mona}
\end{eqnarray}
where \mbox{\boldmath $u$} and \mbox{\boldmath $u'$} are the 
velocities in $K$ and $K'$, 
respectively. We refer now to the propagation of a light pulse whose 
velocity 
\mbox{\boldmath $c_{{\sc k}}$} in $K$ makes an angle $\theta + \pi$
with the $x$-axis (resp. \mbox{\boldmath $c_{{\sc k'}}$} and $
\theta' + \pi$  in $K'$). We have by
projection on the $x-$, $y-$, $x'-$, and $y'-$axis, respectively:
\andy{vero}
\beq
u_x = - c_{{\sc k}}\cos\theta;\;\;\;\;
u_y = - c_{{\sc k}}\sin\theta;\;\;\;\;
u'_x = - c_{{\sc k'}} \cos\theta';\;\;\;\;
u'_y = - c_{{\sc k'}} \sin\theta'
\label{eq:vero}
\eeq
Replacing (\ref{eq:vero}) in (\ref{eq:mona}) and dividing side by side 
leads to 
the aberration formula:
\andy{francesca}
\beq
\tan\theta' =\frac{R(\beta')}{R(\beta)} \;\;
\frac{\sin\theta}{\cos\theta +\frac{\left(\beta'-\beta\right)\left(
1-\beta\cos\theta\right)}{R^2(\beta)}}
\label{eq:francesca}
\eeq
where we have used the fact that in this case 
$c_{{\sc k}}=\frac{c}{1-\beta\cos\theta}$.
When $\beta=0$, (\ref{eq:francesca}) reduces to an equation similar to
(\ref{eq:pascale})
\andy{camille}
\beq
\tan\theta' = R(\beta')\;\; 
                     \frac{\sin\theta}{\cos\theta + \beta'}   
\label{eq:camille}
\eeq 
but there is a great conceptual difference between (\ref{eq:pascale}) and
(\ref{eq:camille}) since in (\ref{eq:camille}), 
$\beta' c$ is a velocity relative to the priviledged frame and 
in (\ref{eq:pascale}), $\tilde{\beta}_1 c$ is the velocity 
between two arbitrarly choosen inertial frames.

The physical basis of aberration is the fact that the velocity of light is
finite and changes its direction when seen from another reference frame. It
is a consequence of the velocity addition formula applied to a light ray
when the {\em observer} is changing its reference frame. There is also an
``aberration effect'' at emission. When the source emits in all directions
it cannot be observed, but when the source emits in only one direction, a
non-uniform motion of the source clearly changes the direction of emission,
since the velocity addition formula can also be applied. Corrections for 
the ``aberration'' of the source are used in the determination of the true
pulsar period in binary systems\cite{data:92a}.

Note that in both theories, only the difference $\Delta\theta$ is 
observable.

\section{Aberration and delay effect}
From (\ref{eq:camille}) one could think that in ether-theories
 the aberration $\theta' - \theta$ due
to an uniform motion relative to $K_0$ should be observable. In the SRT such
a problem does not occur since there is no privileged frame.
 Prokhovnik and
Morris\cite{prmo:89a} wrote: ``Certainly, an observer travelling at very 
great 
speed (say c/2) relative to the universe would have a very distorded and 
asymmetric view of the distribution of the stars and galaxies as a result
of the aberration effect.'' In the same article, they explained that, for 
nearby sources the aberration is cancelled by a delay effect.
In this section, we show that the cancellation of aberration by a delay effect
as nothing to do with the nearness of the source but only with the uniformity
of the motion of the observer, so that an observer travelling at very great
uniform
speed would see no distortion of the distribution of the stars. On the 
contrary,
an observer in a laboratory on earth would see the beam of a laser 
moving of 41'' in six month since the motion of the earth is not uniform.

Let us imagine the following situation: a detector D and a source S emitting
in all directions are at rest in the reference frame $K$ of section 3.
For an observer in $K_0$, the following quantities are defined
(see figure 1):
$l_0$ is the distance from S to the $x_0$-axis, $L_0$ is the projection of
DS on the $x_0$-axis, $\tan\theta_0 =\frac{l_0}{L_0}$. $\theta_0$ is known
by measuring $L_0$ and $l_0$. One could think that by measuring the $\theta'$
of (\ref{eq:camille}), it is possible to determine the velocity of $K$ relative
to $K_0$. It is not so because of a delay effect.
 At time $t_0 = 0$
of $K_0$ the source S begin to emit. The detector D will have moved a
distance $\Delta L_0$ before light from S reaches it. The time $t_0$ of  
reception is
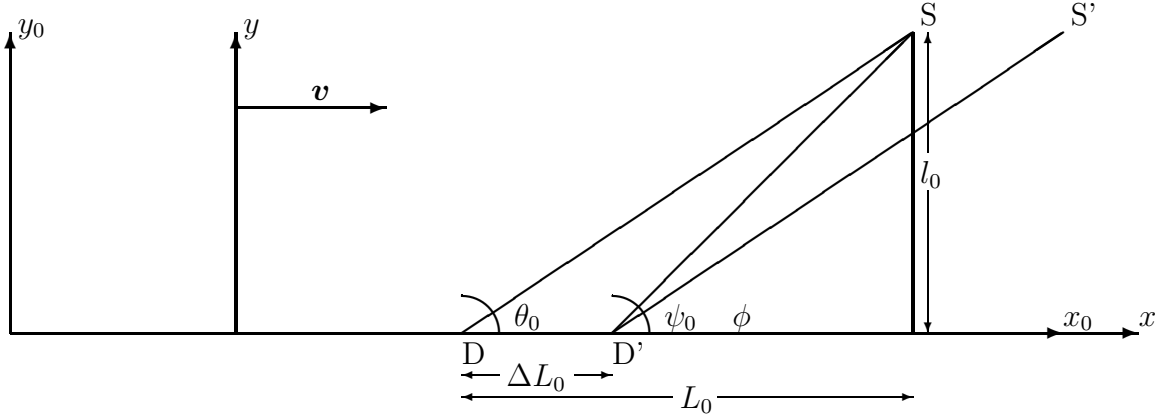
\begin{figure}
\setlength{\unitlength}{1cm}
\begin{picture}(15,5)(0,0)
\thicklines
\put(0,1){\vector(0,1){4}}
\put(0,1){\vector(1,0){14}}
\put(3,1){\vector(0,1){4}}
\put(3,1){\vector(1,0){12}}
\put(3,4){\vector(1,0){2}}
\put(4,4.1){\mbox{\boldmath $v$}}
\put(6,1){\line(3,2){6}}
\put(8,1){\line(1,1){4}}
\put(12,1){\line(0,1){4}}
\put(8,1){\line(3,2){6}}
\put(12.1,5.1){S}
\put(14.1,5.1){S'}
\put(6,0.6){D}
\put(6,1){\oval(1,1)[tr]}
\put(8,0.6){D'}
\put(8,1){\oval(1,1)[tr]}
\put(6.6,0.3){$\Delta L_0$}
\put(8.9,0){$L_0$}
\put(12.1,3){$l_0$}
\put(0.1,5){$y_0$}
\put(3.1,5){$y$}
\put(15,1.1){$x$}
\put(14,1.1){$x_0$}
\put(6.7,1.1){$\theta_0$}
\put(8.7,1.1){$\psi_0$}
\put(9.6,1.1){$\phi$}
\thinlines
\put(6.5,0.45){\vector(-1,0){0.5}}
\put(7.5,0.45){\vector(1,0){0.5}}
\put(8.8,0.15){\vector(-1,0){2.8}}
\put(9.4,0.15){\vector(1,0){2.6}}
\put(12.2,2.9){\vector(0,-1){1.9}}
\put(12.2,3.4){\vector(0,1){1.6}}
\end{picture}
\caption{The source S begins to emit at $t_0 = 0$ but the detector in
D at $t_0 = 0$ recieves the signal only later in D'
 and with an angle $\psi_0$ (in $K_0$). 
Since the detector is at rest in $K$, $\psi_0$ will be aberrated and seen 
as $\phi$ in $K$. An observer in $K$ finds that $\phi = \theta$}
\end{figure}
\andy{suzanne}
\beq
t_0 = \frac{\sqrt{l_0^2 + (L_0 - \Delta L_0)^2}}{c} =\frac{\Delta L_0}{v}
\label{eq:suzanne}
\eeq
so that,
\andy{ruth}
\beq
\Delta L_0 = \frac{\beta}{R^2(\beta)}\;\;\left(-L_0\beta \; +\; \sqrt{\left(
R(\beta)l_0\right)^2 + L_0^2}\right)
\label{eq:ruth}
\eeq
The angle $\psi_0$ due to a delay effect  is the angle at which light falls
on D in $K_0$ and can now be calculated as:
\andy{fabienne}
\beq
\tan\psi_0 = \frac{l_0}{L_0\left( 1 - \frac{\Delta L_0}{L_0}\right)} =
\frac{\sin\theta}{R(\beta)\left[\cos\theta-\frac{\beta\left(
1-\beta\cos\theta\right)}{R^2(\beta)}\right]}
\label{eq:fabienne}
\eeq
where we have used $\theta$, the angle in $K$ corresponding to $\theta_0$ 
given by the transformations:
\andy{moon}
\beq
L = \frac{L_0}{R(\beta)};\;\;\; \Delta L = \frac{\Delta L_0}{R(\beta)}
;\;\;\;\; l=l_0;\;\;\;\;
\tan\theta=R(\beta)\tan\theta_0.
\label{eq:moon}
\eeq
The angle $\psi_0$ is seen as $\phi$ in $K$. We have the following aberration
relation. 
\andy{sandra}
\beq
\tan\psi_0 = 
\frac{\sin\phi}{R(\beta)\left[\cos\phi-\frac{\beta\left(
1-\beta\cos\phi\right)}{R^2(\beta)}\right]}
\label{eq:sandra}
\eeq
The relation (\ref{eq:sandra}) is given here for the need of the proof in its
inverse form and was obtained by putting $\beta' = 0$, $\psi_0$ in place of
$\theta'$ and $\phi$ in place of $\theta$ in (\ref{eq:francesca}).
Let us write relation (\ref{eq:fabienne}) as $\psi_0 = {\cal F}(\theta)$
and relation (\ref{eq:sandra}) as $\psi_0 = {\cal G}^{-1}(\phi)$
so that we obtain:
\andy{kirst}
\beq
\phi = {\cal G}(\psi_0) = {\cal G}({\cal F}(\theta)) = \theta
\label{eq:kirst}
\eeq
since ${\cal F} = {\cal G}^{-1}$.
It means that an observer in $K$ measures an angle $\phi$ exactly equal to
$\theta = \arctan\left(\frac{l}{L}\right)$. So in the TIT, the aberration 
due to a uniform
motion is not observable. No assumptions was made concerning $l_0$ and $L_0$, 
so that
the result obtained here is exactly the same if the source is near to or far 
from the detector solong the motion is uniform.
 
\section{Gedankenexperiment}
Aberration of light is typically a physical phenomenon in which the one-way
velocity of light comes into play.
Sj\"{o}din and Podlaha\cite{sjpo:81a} wrote about stellar aberration: 
``The {\em only} cause
of the effect is the change of the relative velocity of the earth during the 
year.(...).The use of so-called {\em absolute synchronisation} could, 
therefore, 
impossibily have an influence on the observed effect.'' The point is that
Sj\"{o}din and Podlaha used in their proof only one inertial frame 
representing the 
earth, thus a frame which is totally unable to represent the change of the 
velocity of the earth during the year. As a better approximation to reality 
we propose here a Gedankenexperiment with two different reference frames. 

Let us begin with the SRT. A spaceship is at rest in an inertial system $K$.
On the inside, one prepares a laser so that its beam makes an angle  
$\theta$ with the wall of the rocket which is also the future direction
of acceleration. Then the rocket accelerates and reaches a velocity 
\mbox{\boldmath $\tilde{v}_1$} relative to $K$ (see figure~2). Of course 
in $K$, clocks are synchronized with Einstein's procedure, and $\tilde{v}_1$
is measured with such clocks.
\begin{figure}
\setlength{\unitlength}{1cm}
\begin{picture}(15,6)(0,0)
\thicklines
\put(0.5,1){\framebox(4,3)}
\put(2.5,4.1){\bf L}
\put(2.5,0.1){$K_0$}
\put(4.6,1.1){$x_0$}
\put(0.1,5){$y_0$}
\put(2.5,4){\line(-1,-5){0.6}}
\put(2.3,1.2){$\theta_0$}
\put(1.9,1){\oval(0.8,0.8)[tr]}
\put(5.5,1){\framebox(4,3)}
\put(7.5,4.1){\bf L}
\put(7.5,0.1){$K$}
\put(9.6,1.1){$x$}
\put(5.1,5){$y$}
\put(7.5,4){\line(-1,-3){1}}
\put(6.9,1.2){$\theta$}
\put(6.5,1){\oval(0.8,0.8)[tr]}
\put(10.5,1){\framebox(4,3)}
\put(12.5,4.1){\bf L}
\put(14.6,1.1){$x'$}
\put(10.1,5){$y'$}
\put(12.5,0.1){$K'$}
\put(12.5,4){\line(-1,-2){1.5}}
\put(11.4,1.2){$\theta'$}
\put(11,1){\oval(0.8,0.8)[tr]}
\put(0,1){\vector(1,0){4.75}}
\put(5,1){\vector(1,0){4.75}}
\put(10,1){\vector(1,0){4.75}}
\put(0,1){\vector(0,1){4}}
\put(5,1){\vector(0,1){4}}
\put(10,1){\vector(0,1){4}}
\end{picture}
\caption{Different angles laser beam/$x$-axis. 1. In the TIT: $K_0$
(privileged frame), $K$(initial inertial frame; 
``absolute'' velocity: \mbox{\boldmath $v$}), and 
$K'$(inertial frame after acceleration; ``absolute'' velocity: 
\mbox{\boldmath $v'$}; velocity relative to $K$: \mbox{\boldmath $v_{1}$}).
2. In the SRT: $K$(initial inertial frame), $K'$(inertial frame after 
acceleration; velocity relative to $K$: \mbox{\boldmath $\tilde{v}_1$}. 
L stands for laser source.}
\end{figure}
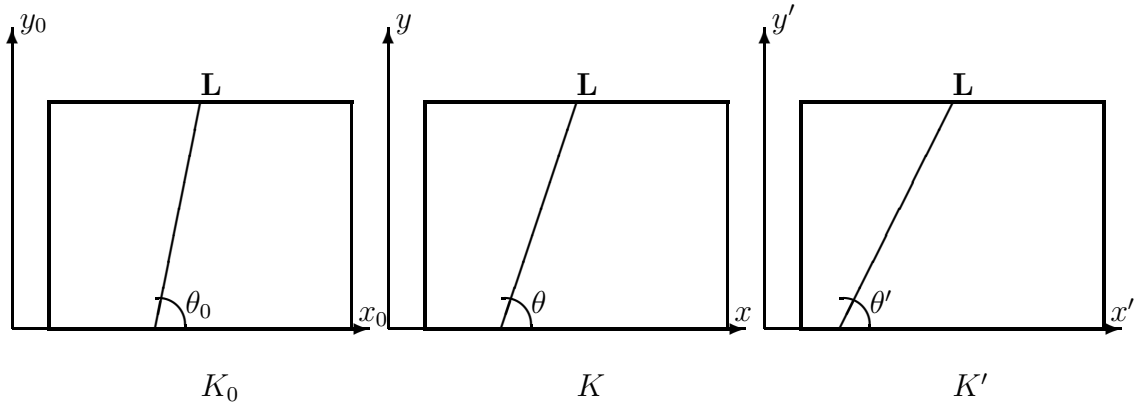 
The angle $\tilde{\theta}'$ after acceleration is 
given by (\ref{eq:pascale}). 

Let us now consider the same experiment in the TIT, but imagine that $K$ 
was already moving with a velocity \mbox{\boldmath $v$} 
relative to $K_0$. For simplicity, we suppose that the rocket reaches a
velocity \mbox{\boldmath $v_1$} relative to $K$ which is parallel to
\mbox{\boldmath $v$} and has the velocity \mbox{\boldmath $v'$} 
relative to $K_0$. In the TIT, clocks are synchronized with an ``absolute'' 
procedure and  $v_1$ is measured with these clocks. $v'$ is given by
the first part of (\ref{eq:mona}), with $u'_x = 0$ and $u_x = v_1$ , 
respectively . Since we want to predict aberration in the 
same physical situation as above, and also be able to compare the results with
those of the SRT, we express $v_1$ in terms of $\tilde{v}_1$ and $v$ by
means of (\ref{eq:genevieve}). We obtain:
\andy{emmanuelle}
\beq
v' = v+\frac{\tilde{v}_1}{1+\beta\tilde{\beta}_1}R^2(\beta)
\label{eq:emmanuelle}
\eeq
The angle $\theta'$ is now given by (\ref{eq:francesca}) which seems to be very 
different from (\ref{eq:pascale}) and in particular to depend on $v$. 
It is in fact not the case. Replacing $v'$ in (\ref{eq:francesca}) by its 
value (\ref{eq:emmanuelle}), one obtains after a few calculations that 
{\bf(\ref{eq:francesca}) is exactly the same as (\ref{eq:pascale})}. So 
there is
no observable difference between the SRT and the TIT in this case.

\section{Discussion}
\begin{enumerate}
\item The arguments of sections 2, 4, and 5
 bring strong evidences that also in other cases where one could try to 
obtain a difference between the TIT and the SRT by comparing an observed effect
in two different inertial  frames, one would also obtain the same conclusion 
as here.
\item Nevertheless, the conclusion we obtain here cannot be generalised further
without proof. We cannot conclude that there is a general equivalence between
the SRT and the TIT.
\item It is still an open question to know if the one way velocity of light
is a purely conventional quantity or is fixed by nature itself. For the time
being it seems that the TIT and the SRT are equivalent for all observational
purpose but there is a great philosophical difference between them.
\end{enumerate}

\section{Conclusion}
By means of a Gedankenexperiment, we have proved that the SRT and the TIT
lead to the same results in the case of aberration of light. In spite of 
the fact that hundreds of experiments trying to detect an ether drift have
obtained a negative result no general proof of the equivalence of theories
built on different synchronizations was ever brought. The question of
the conventionality of the one-way velocity of light is still open.

\section{Acknowledgement}
I want to thank Prof. Franco Selleri for his suggestions and criticisms
and Gino Lepore for his precious help.

\end{document}